\documentclass[12pt,preprint]{aastex}

\usepackage[font=small]{caption}
\slugcomment{Accepted by ApJ November 8 2013}

\shorttitle{Resolved Spectroscopy of WISEPC J121756.91+162640.2AB}
\shortauthors{Leggett et al.}

\begin{document}

%% LaTeX will automatically break titles if they run longer than
%% one line. However, you may use \\ to force a line break if
%% you desire.

%%\title{Near-Infrared Photometric Followup of WISE Y Dwarfs}
\title{Resolved Spectroscopy of the T8.5 and Y0--0.5 Binary \\ WISEPC J121756.91+162640.2AB}

%% Use \author, \affil, and the \and command to format
%% author and affiliation information.
%% Note that \email has replaced the old \authoremail command
%% from AASTeX v4.0. You can use \email to mark an email address
%% anywhere in the paper, not just in the front matter.
%% As in the title, you can use \\ to force line breaks.

\author{S. K. Leggett\altaffilmark{1}}
\email{sleggett@gemini.edu}
\author{Michael C. Liu\altaffilmark{2}}
\author{Trent J. Dupuy\altaffilmark{3}}
\author{Caroline V. Morley\altaffilmark{4}}
\author{M. S. Marley\altaffilmark{5}}
\and
\author{D. Saumon\altaffilmark{6}}

\altaffiltext{1}{Gemini Observatory, Northern Operations Center, 670
  N. A'ohoku Place, Hilo, HI 96720, USA} 
\altaffiltext{2}{Institute for Astronomy, University of Hawaii, 2680 Woodlawn Drive, Honolulu, HI 96822, USA}
\altaffiltext{3}{Smithsonian Astrophysical Observatory, 60 Garden Street, MS 9, Cambridge, MA 02138, USA}
\altaffiltext{4}{Department of Astronomy and Astrophysics, University of California,
Santa Cruz, CA 95064, USA}
\altaffiltext{5}{NASA Ames Research Center, Mail Stop 245-3, Moffett Field, CA 94035, USA}
\altaffiltext{6}{Los Alamos National Laboratory, PO Box 1663, MS F663, Los Alamos, NM 87545, USA}

\begin{abstract}
We present 0.9 -- 2.5 $\mu$m resolved spectra for  the ultracool binary  WISEPC J121756.91+162640.2AB. The system consists of a pair of brown dwarfs that straddles the currently defined T/Y spectral type boundary. We use synthetic spectra generated by model atmospheres that include chloride and sulfide clouds (Morley et al.), the distance to the system (Dupuy \& Kraus), and the radius of each component based on evolutionary models (Saumon \& Marley) to determine a probable range of physical properties for the binary.  The effective temperature of the T8.5 primary is 550 -- 600~K, and that of the Y0 -- Y0.5 secondary is $\approx\,$450~K.   The atmospheres of both components are either free of clouds or have extremely thin cloud layers.
We find that the masses of the primary and secondary are 30 and 22  $M_{\rm Jup}$, respectively, and that the age of the system is 4 -- 8 Gyr. This age is consistent with astrometric measurements (Dupuy \& Kraus) that show that the system has kinematics intermediate between those of the thin and thick disks of the Galaxy. An older age is also consistent with an indication by the $H - K$ colors that the system is slightly metal-poor.
\end{abstract}


%% From the front matter, we move on to the body of the paper.
%% In the first two sections, notice the use of the natbib \citep
%% and \citet commands to identify citations.  The citations are
%% tied to the reference list via symbolic KEYs. The KEY corresponds
%% to the KEY in the \bibitem in the reference list below. We have
%% chosen the first three characters of the first author's name plus
%% the last two numeral of the year of publication as our KEY for
%% each reference.

\section{Introduction}

The   Wide-field Infrared Survey Explorer ({\it WISE}, Wright et al. 2010) has significantly advanced the study of brown dwarfs, stellar-like objects that have insufficient mass for stable hydrogen burning (Kumar 1963, Hayashi \& Nakano 1963). About the size of Jupiter, the cool brown dwarfs are intrinsically very faint. Prior to {\it WISE}, 
brown dwarfs as cool as $T_{\rm eff} = 500$~K  had been found  in near-infrared surveys undertaken by  4$\,$m-class ground-based telescopes (e.g. Lucas et al. 2010). Two brown dwarfs were also known with   $T_{\rm eff} \approx 400$~K: CFBDSIR J145829+101343B, a companion to a warmer brown dwarf, discovered using laser guide star adaptive optics imaging (Liu et al. 2011), and GJ 3483B, a companion to a white dwarf, found using the Infrared Array Camera on the {\it Spitzer Telescope} in a proper motion search for faint companions (Luhman, Burgasser \& Bochanski 2011). These 400~K brown dwarfs were expected to have a later spectral type than all previously known T dwarfs. Observations  beyond the near-infrared, such as those obtained with the {\it Spitzer Telescope}, are advantageous for studying  cool brown dwarfs because 
objects with  $T_{\rm eff} < 700$~K emit  $> 50$\% of their energy at $\lambda > 3$ $\mu$m (Leggett et al. 2010a). In 2011  the mid-infrared 0.4$\,$m  {\it WISE} telescope  
 identified the first
large sample of brown dwarfs with  $T_{\rm eff} < 500$~K
(Cushing et al. 2011, Kirkpatrick et al. 2011 and 2012).
Kirkpatrick et al. (2011) presented the first hundred {\it WISE}  brown dwarfs and this paper studies one of those objects, WISEPC J121756.91+162640.2 (hereafter WISE 1217$+$16), which was classified as a T9 spectral type by Kirkpatrick et al.

WISE 1217$+$16 was found to be a binary system by Liu et al. (2012, hereafter Liu12) using Keck laser guide star adaptive optics. The pair is separated by $0\farcs 76$ at a position angle of 14.3$^\circ$ .
It is an unusual binary, having a relatively wide separation and a large difference in near-infrared brightness between the two components (see Table 1). Liu12 obtained resolved $H$-band spectra of the two components and classified WISE 1217$+$16A as a T9 and WISE 1217$+$16B as a Y0, using the spectral classification scheme for the latest T-type and early Y-type brown dwarfs proposed by Cushing et al. (2011). Here we present resolved spectra for the system covering a wider wavelength range of 0.9 -- 2.5 $\mu$m and expand on the analysis presented by  Liu12.

\section{Observations}

We used the Gemini near-infrared spectrograph (GNIRS, Elias et al. 2006)  to obtain  0.9 -- 2.5 $\mu$m cross-dispersed spectra of  WISE 1217$+$16AB via queue program GN-2012B-Q-28. In order to resolve the pair, observations were carried out only when the natural seeing full width half maximum (FWHM) was $0\farcs 45$ or better and only at airmasses less than 1.2. Both photometric and cloudy conditions were utilised. The $0\farcs 3$ slit was used with the $0\farcs 15$ pixel$^{-1}$ camera, resulting in a resolving power $\lambda/\delta\lambda \sim 1700$. The slit was placed at 14.3$^\circ$ so that both sources were in the slit. Individual exposure times of 300 seconds were used, with the target nodded along the slit. The A0 star HD 101060 and the F4 star HD 114072 were used as calibrators to remove telluric features.

Observations were obtained on 2013 February 21, April 9, April 26 and May 8 UT. Time on source on these nights was 35, 40, 60 and 60 minutes, respectively. The data from  2013 February 21 were obtained in thicker cloud cover, and as the signal was around half that of the three other nights those data were omitted.
Calibration lamps on the telescope provided data for wavelength calibration and flat fielding, as well as pinhole images for tracing the cross-dispersed data along the detector.  Figure 1 shows the flat fielded, sky-subtracted and rectified  $J$-band spectrum from 2013 May 8 as a two-dimensional image.  The  FWHM is around 2 pixels ($0\farcs 3$) and the separation between components is  5 pixels ($0\farcs 76$).

The spectrum of both the A and B components were extracted using apertures centered at the respective peaks, with a lower limit of $-2.5$ pixels and an upper limit of 2.5 pixels. Where B is extremely faint, we used the known offset from A to place the aperture. The contribution of the brighter component to the fainter component's spectrum was determined by extracting the signal of the A component at the location of the B components aperture on the opposite side of the profile. Typically the contribution was 5\% of the signal of the A component, and 20\% of that of the B component. To avoid adding noise, we subtracted an appropriately scaled version of the   A spectrum from the B spectrum, and not the spectrum extracted from the wing. The spectral orders for each component were combined, averaging the regions of overlap at 0.98 -- 0.95, 0.98 -- 1.06, 1.13 -- 1.25, 1.45 -- 1.52, 1.88 -- 1.90 $\mu$m.  Each spectrum was flux-calibrated using the photometry presented in Liu12. The final  spectrum for each component was constructed by averaging each night's data. Based on the scatter between the three measurements, the uncertainty in the flux for the A component is 7\% over the $Y$-band peak, 5\% over the $J$-band peak, 7\% over  the $H$-band peak, and 15\% over the $K$-band peak. Similarly for the B component  the uncertainty in the flux is 20\% over the $Y$-band peak, 15\% over the $J$-band peak, 10\% over  the $H$-band peak, and 60\% over the $K$-band peak (where there is very little flux).
 
Figure 2 shows the spectrum of each component derived here, together with a comparison of the summed spectrum to the spectrum of the unresolved pair presented by Kirkpatrick et al. (2011). The inset plot compares our $H$-band sections to the resolved spectra obtained previously at Keck by Liu12. The agreement with the Keck data is excellent, and that with the unresolved spectrum is reasonable, given the noise in each measurement. These comparisons validate our extraction of the one-dimensional spectra from the two-dimensional images.

\section{Spectral Classification}

Figure 3 compares our spectrum of  WISE 1217$+$16A and  WISE 1217$+$16B to T and Y dwarfs classified by Cushing et al. (2011) as T8.5, T9, Y0 and Y0.5. The comparison spectra have been scaled to the $J$-band flux peak of each component. Inset plots zoom in on the $J$-band flux peak, the width of which is a diagnostic of the spectral type (Warren et al. 2007, Burningham et al. 2008, Cushing et al. 2011, Kirkpatrick et al. 2012, Mace et al. 2013). It is clear visually that   WISE 1217$+$16A is an excellent match to the T8.5 reference, and has a wider  $J$-band flux peak than the T9. However  WISE 1217$+$16A is much fainter at  $K$ than either reference source, which we discuss in the following paragraph. We classify  WISE 1217$+$16A as  T8.5.

The  WISE 1217$+$16B comparison is less straightforward. The blue side of the $J$-band flux peak matches the Y0.5 template, however the red side is in better agreement with the Y0 template. Both the blue and red sides of the $J$-band peak are impacted by absorption by CH$_4$, H$_2$O and NH$_3$; the blue side is also affected by pressure-induced H$_2$ absorption (e.g. Figure 1 of Leggett et al. 2009). It is likely that metallicity and gravity, as well as temperature, impact the shape  of the $J$-band  peak and that we are  seeing variations in these parameters for the three Y dwarfs. The fact that the $K$-band is suppressed for   WISE 1217$+$16A implies that it either has a high gravity or low metallicity which enhances the H$_2$ absorption at 2 $\mu$m (e.g. Leggett et al. 2009); possibly we are also seeing enhanced H$_2$ absorption  in the blue wing of the  $J$-band  peak. We define  WISE 1217$+$16B as Y0 -- 0.5.

Note that the near-infrared spectra of these components, which straddle the current T/Y classification boundary, are overall very similar. Increased molecular absorption narrows the flux peaks at lower $T_{\rm eff}$, but there is otherwise no strong spectral marker for the transition from T to Y
in the near infrared. One difference that is apparent is the height and width of the $Y$-band peak around 1 $\mu$m --- relative to the  $J$-band peak, the later spectral type and cooler object has a taller and broader  $Y$-band peak. This is also seen in the $Y - J$ colors, which become bluer. Figure 4 is the  {$Y - J$,$M_Y$}  color-magnitude diagram, showing the trend to bluer  $Y - J$. Also, this Figure shows about half the drop in absolute magnitude between T9 and Y0 than is seen for 
$M_J$ and $M_H$ (Dupuy \& Kraus 2013, Figure 5 and \S 4.3). For T9 types with $T_{\rm eff} \approx 500$~K, the alkali elements are condensing into chlorides and sulfides, weakening the strong $0.77 \mu$m K I resonance doublet, the red wing of which suppresses the $Y$-band flux. This likely explains the brightening at $Y$ (Liu12, Leggett et al. 2012). 
There is a lot of scatter in the $Y - J$ colors of early Y dwarfs (Figure 4), again suggesting that there are variations in metallicity and gravity, as well as temperature, within this small sample. This disparity can also be seen spectroscopically in Figure 3:  WISE 1217$+$16B has a broader $Y$-band peak than the comparison Y dwarfs and is bluer in $Y - J$. 

The gap in $M_Y$ and  $M_H$ at the T to Y transition is striking in Figures 4 and 5. This gap is not seen at mid-infrared wavelengths (Leggett et al 2013, hereafter Leg13; Dupuy \& Kraus 2013). The gap may disappear as more very late-type T and Y dwarfs are found and more distances are determined but it is tempting to associate the distinct drop in flux at 1 -- 2 $\mu$m at  $T_{\rm eff} \approx 400$~K with a physical phenomenon. Possibilities include  the appearance of water clouds or the impact of a newly significant opacity such as H$_2$. Improved models are required to explore this further.

Finally we note that the object initially identified as the prototype Y dwarf, WISEP J182831.08+265037.8     (Cushing et al. 2011), appears to be different from the other Y dwarfs in terms of colors and luminosity (\S 4.3, Beichman et al. 2013, Dupuy \& Kraus 2013, Leg13, Kirkpatrick et al. 2013). Once a larger sample of Y dwarfs is known it is likely that the spectral classification scheme will have to be revisited. Also as the models are improved with more complete treatment of opacities, clouds and turbulence, we hope to disentangle the effects of temperature, metallicity and gravity on the near-infrared spectrum of these cold objects.

\section{Comparison to the Models}

\subsection{The Models}

Leg13 compares near-infrared photometry to cloud-free models and to a
new generation of models that include clouds consisting of sulfide and chloride condensates; we use the same models here. 
The cloud-free model atmospheres are as described in Saumon \& Marley (2008) and Marley et al. (2002), with updates to the line list of NH$_3$ and of the collision-induced absorption of H$_2$ as described in Saumon et al. (2012). 
The cloudy models are described in detail in Morley et al. (2012).
Morley et al. have added absorption and scattering by condensates of Cr, MnS, Na$_2$S, ZnS and KCl to the cloud-free models. These condensates  have been predicted to be present in low-$T_{\rm eff}$ atmospheres by Lodders (1999) and  Visscher, Lodders \& Fegley (2006). Morley et al.  use the Ackerman \& Marley (2001) cloud model to account for these previously neglected clouds. The vertical cloud extent is determined by balancing upward turbulent mixing and downward sedimentation.  A parameter $f_{\rm sed}$ 
describes the efficiency of sedimentation, and is the ratio of the sedimentation velocity to the convective velocity; lower values of $f_{\rm sed}$ imply thicker  (i.e. more vertically extended) clouds.  We have found that models
that include iron and silicate grains and  which have $f_{\rm sed}$ of typically 2 -- 3 fit L dwarf spectra well, 
those with  $f_{\rm sed}$ 2 -- 4 fit T0 to T3 spectral types well, and  cloud-free models fit T4 -- T8 types well 
(e.g.  Saumon \& Marley 2008, Stephens et al. 2009). However for the latest T-types significant discrepancies exist between the models and the observations, in the near-infrared (e.g. Leggett et al. 2009, 2012). The new models with the chloride and sulfide clouds help to resolve these discrepancies, because these clouds are significant for 
dwarfs with $T_{\rm eff} = 400$ -- 900~K (approximately T7 to Y1 spectral types), with a peak impact at around 600~K. Below $T_{\rm eff} \sim 400\,\rm K$ water clouds are expected to 
form (Burrows, Sudarsky \& Lunine 2003), which have not yet been incorporated into the models (although water condensation is accounted for in the gas opacity). 

The models used in the present analysis have solar metallicity and neglect  departures from chemical equilibrium caused by vertical mixing.
The mixing enhances the abundance of CO and CO$_2$ and reduces the 5 $\mu$m flux (Saumon et al. 2006).  Vertical mixing also decreases the abundance of NH$_3$, which would otherwise produce stronger absorption features at 1.03 and 1.52 $\mu$m  than are seen in the known Y dwarfs (Leg13). The mixing can be parameterized with the eddy diffusion coefficient $K_{zz}$ cm$^2$ s$^{-1}$, where
values of log $K_{zz} = 2$ -- 6  corresponding to mixing timescales of $\sim 10$ yr to $\sim 1$ hr, respectively, reproduce the observations of T dwarfs (e.g. Saumon et al. 2007). Leggett et al. (2012) find that the  
$T_{\rm eff}= 500\,$~K dwarf UGPS J0722$-$0540 is undergoing vigorous mixing, with log $K_{zz} \approx 5.5$ -- 6.0, and this impacts the  {\it WISE}  4.6 $\mu$m W2 band by $\gtrsim 0.3$ magnitude. Leg13 find that increasing the calculated W2 flux by 0.3 magnitude results in the model sequences reproducing the observed color trends in T and Y dwarfs quite well.

\subsection{Previously Determined Properties of WISE 1217$+$16AB}

Liu12 derived a photometric distance of 10.5 $\pm 1.7$ pc for  WISE 1217$+$16A based on a spectral type assignment of T9 and   a $J$-band bolometric correction  for very late-type T dwarfs. The luminosity was combined with evolutionary models to derive physical properties for ages of 1 Gyr and 5 Gyr. For the younger age the inferred  $T_{\rm eff}$ and mass are 550~K and 13 $M_{\rm Jup}$ for the primary, and  400~K and 7 $M_{\rm Jup}$ for the secondary. For the older age these values are 650~K and 33 $M_{\rm Jup}$ for the primary, and  400~K and 17 $M_{\rm Jup}$ for the secondary. Dupuy \& Kraus (2013) have recently published a trigonometric distance to the binary of  $10.1_{-1.4}^{+1.9}\,$pc. Using this distance and model-based bolometric corrections to the summed observed flux given by the measured magnitudes, they determine, for an age of 5 Gyr,   $T_{\rm eff}$ and mass of  600~K and 31 $M_{\rm Jup}$ for the primary, and  450~K and 19 $M_{\rm Jup}$ for the secondary.

Leg13 compare the observed resolved near-infrared colors of the WISE 1217$+$16 components to the Morley et al. (2012) cloudy models. The W2 magnitudes for each component are estimated based on spectral type, constrained by the unresolved W2 value. Leg13 find that the  model sequences are consistent with a single-age solution for the binary. Higher gravity solutions, corresponding to an age $\sim$5 Gyr, provided better fits than the lower gravity corresponding to  1 Gyr, because of the relatively blue $H - K$ color. The colors also suggested that the atmospheres of both components had thin cloud layers with  $f_{\rm sed} \approx 5$.

The proper motion for the binary is measured to be $1\farcs45 \pm 0\farcs04$ yr$^{-1}$ (Dupuy \& Kraus 2013), implying a tangential velocity of $70 \pm 10$ km s$^{-1}$. This velocity implies kinematics intermediate between the thin and thick disk  populations (e.g. Brook et al. 2012; Dupuy \& Liu 2012) and therefore an age around 7 Gyr.
This result is consistent with the findings of Leg13, that the higher gravity and therefore older age of 5 Gyr is favored over the younger 1 Gyr age.

\subsection{Color-Magnitude and Spectral Energy Distribution}

Figure 5 illustrates the location of  WISE 1217$+$16A and  WISE 1217$+$16B in a near-infrared color-magnitude diagram. 
Here $M_H$ is used as the luminosity indicator as $H$ is less sensitive to the clouds than $Y$ or $J$, and less sensitive to metallicity and gravity than $K$.
The intrinsically faintest sources are identified.
%(1) CFBDSIR J145829+101343A  (Delorme et al. 2010, Liu et al. 2011)
%(2) BD $+01^{\circ} 2920$B (Pinfield et al. 2012) 
%(3) SDSS J141624.08+134826.7B (Burgasser, Looper \& Rayner 2010; Burningham et al. 2010; Scholz 2010) 
%(4) ULAS J133553.45+113005.2 (Burningham et al. 2008)
%(5) Wolf 940B (Burningham et al. 2009) 
%(6) CFBDS J005910.90-011401.3 (Delfosse et al. 2008)
%(7) $\xi$ UMaC (Wright et al. 2013)
%(8) UGPS J072227.51−054031.2 (Lucas et al. 2010)
%(9)  CFBDSIR J145829+101343B  (Delorme et al. 2010, Liu et al. 2011)
%(10)  WISEP J173835.52+273258.9   (Cushing et al. 2011)
%(11) WISEPC J014807.25–720258.8 (Cushing et al. 2011)
% (12) WISEPC J205628.90+145953.3 (Cushing et al. 2011)
%(13) WISEP J154151.65–225025.2 (Cushing et al. 2011)
%(14) WISEPC J140518.40+553421.5 (Cushing et al. 2011)
%(15) WISEP J182831.08+265037.8     (Cushing et al. 2011)
%(16) WISE J035934.06–540154.6 (Kirkpatrick et al. 2012)
%(17) WISE J064723.23−623235.5 (Kirkpatrick et al. 2013).
Photometry and parallaxes are taken from Leg13 and references therein, updated by measurements from Beichman et al. (2013), Dupuy \& Kraus (2013), Kirkpatrick et al. (2013), Mace et al. (2013), Marsh et al. (2013) and Wright et al. (2013). Note that the latest type Y dwarf currently identified,  WISEP J182831.08+265037.8 (\#15), appears to be unusually bright in $H$  (Beichman et al. 2013, Dupuy \& Kraus 2013; see also \S 3).

The typical spectral types and approximate $T_{\rm eff}$ at particular $M_H$ are shown along the right axis of Figure 5
(e.g. Burningham et al. 2010; Dupuy \& Kraus 2013; Leggett et al. 2009, 2010a,b, 2012, 2013; Liu12; Pinfield et al. 2012; Smart et al. 2010; Wright et al. 2013). The  $T_{\rm eff}$ have been derived primarily from luminosity arguments and evolutionary models.
There are three brown dwarfs with  $T_{\rm eff} \approx 600$~K and $\log g \approx 5.0$ that are of particular interest:  
BD $+01^{\circ} 2920$B (2), SDSS J141624.08+134826.7B (3) and Wolf 940B (5). These are companions to a  G, L and M dwarf, respectively. Although  $T_{\rm eff}$ and  $\log g$ are similar,  Wolf 940B has a metallicity close to solar, while  SDSS J141624.08+134826.7B  and   BD $+01^{\circ} 2920$B have [Fe/H] $\lesssim -0.3$ dex (Burgasser et al. 2010, Burningham et al. 2010, Leggett et al. 2010b, Pinfield et al. 2012; see also Burningham et al. 2013). The impact of the lower metallicity is clearly seen in the $H - K$ color in Figure 5, which also suggests that the  WISE 1217$+$16 system may be slightly metal poor. If, instead, the blue $H - K$ is due to gravity only, the size of the shift ($\approx 0.3$ magnitudes) implies a gravity $\sim 1.0$ dex higher than typical for the type (e.g. Burningham et al. 2013, their Figure 10).  A gravity this large is unlikely, given that at  $T_{\rm eff} \approx 600$~K an increase in age from 1 to 10 Gyr corresponds to an increase in gravity of 0.6 dex (Saumon \& Marley 2008, their Figure 4). Hence we suggest that the system has a relatively high gravity combined with a metallicity of about $-0.1$ dex.

It can be seen that the absolute $H$ magnitudes for the  WISE 1217$+$16 components are consistent with the assigned spectral types of T8.5 and Y0--0.5. Using the previous studies of late T and early Y dwarfs, the figure suggests that the components have  $T_{\rm eff}=550$ -- 600$\,$K and 
400 -- 450$\,$K, respectively, which is also consistent with previous determinations (\S 4.2).
Table 2 gives physical parameters for each component for these values of $T_{\rm eff}$, for a range in age of 4 Gyr to 10 Gyr, calculated using the evolutionary models of Saumon \& Marley (2008).

Figure 6 shows the spectrum of each component and synthetic spectra generated by the Morley et al. (2012) models.
The spectra have been scaled by the known distance to the binary (Table 1), and the radius of each component as given by the evolutionary models for an age of 6 Gyr (Table 2). If the system is younger the radius is larger and the synthetic spectrum would be brighter, and vice versa. For ages of 2 and 4 Gyr the scaling factor is 20\% and 8\% larger, while for ages of 8 and 10 Gyr the factor is 6\% and 8\% smaller, respectively. Strictly, if  $T_{\rm eff}$ is kept constant and radius changed then gravity also changes, which would impact the spectral energy distribution. However for the purpose of constraining the model by the luminosity, the gravity impact is small. For example, if $T_{\rm eff} = 500$~K, and age is 6 Gyr, then radius and gravity are $R=0.0938\,R_\odot$ and $\log g {\rm (cgs)}=4.888$. If $T_{\rm eff} = 500\,$~K, and age is 10$\,$Gyr, then radius and gravity are 0.0896$\,R_\odot$ and $\log g=5.026$ (Saumon \& Marley 2008). The spectral change due to a change in gravity of 0.14 dex is small (e.g. Leggett et al. 2009), while the change in the flux scaling factor ($R^2$) of 10\% is significant. 

Another factor to bear in mind when examining Figure 6 is that the models do not include the vertical mixing that likely occurs in such atmospheres (e.g. Leg13, \S 4.1). The mixing is expected to increase the abundance of N$_2$ at the expense of that of NH$_3$. In the near-infrared this 
affects the $H$-band in particular; the NH$_3$ absorption is much reduced, making the blue wing and peak of the $H$-band brighter. The effect is $\sim$ 20\% at the peak of the $H$-band at these temperatures (based on preliminary Saumon \& Marley models). 

The model comparison in Figure 6 and the relative strengths of the $Y$, $J$ and $H$ peaks ($K$ is very faint) suggests
that each component of the WISE 1217$+$16 binary has very thin to no sulfide/chloride clouds: $f_{\rm sed} \gtrsim 6$.
The brightness of the flux peaks, especially considering that the $H$ peak is likely to be brighter when mixing is included, suggests that the age is unlikely to be less than 4 Gyr, as the model spectra will then be too bright. The system may be as old as 10 Gyr, as shown in the inset in Figure 6, although there is a discrepancy at $J$ for the B component in that case. Enhanced H$_2$ absorption, if the system is metal poor, could reduce the flux at $K$, and possibly at the blue wing of $J$ for the cooler dwarf, improving the fit. The relative height of the flux peaks are likely to also be sensitive to the detailed structure of any cloud decks, as is seen at the L/T dwarf transition (e.g. Marley et al. 2012, Apai et al. 2013).
In summary, plausible fits are obtained for an age range of 4 to 8 Gyr. 

For the A component the best match to the models occurs if  $550 \lesssim T_{\rm eff}$~(K) $\leq 600$ and there are no clouds.  For the B component the best match to the models occurs if  $T_{\rm eff} \approx 450$~K  and there are no clouds or an extremely thin sulfide/chloride cloud layer with $f_{\rm sed} > 5$. Other spectral comparisons (not shown), where the synthetic spectra are scaled by the known distance and the evolutionary-determined radius, showed that we can exclude  $T_{\rm eff}$ values of 500~K for either the A or B component, due to large discrepancies in brightness levels. Similarly temperatures as high as 650~K can be excluded for WISE 1217$+$16A. Although we do not have 350~K model spectra, luminosity arguments (see Figure 5 and Dupuy \& Kraus 2013, their Table S5) show that  WISE 1217$+$16B cannot be as cool as 350~K.

Table 3 summarises the likely values for the physical properties for the system.

\section{Conclusion}

We have used 
nights of excellent seeing on Mauna Kea to  obtain resolved 0.9 -- 2.3 $\mu$m spectra for the T dwarf and Y dwarf binary  WISEPC J121756.91$+$162640.2AB. The spectral extraction is confirmed to be accurate by comparison to the unresolved spectrum and resolved $H$-band spectra obtained previously. Comparison to the near-infrared spectra of (the small number of) very late-type T dwarfs and early-type Y dwarfs implies spectral types of T8.5 and Y0 -- Y0.5 for the primary and secondary, respectively. Thus the system  straddles the currently defined T/Y spectral type boundary.

Using synthetic spectra generated by model atmospheres that include chloride and sulfide clouds (Morley et al. 2012), and constrained by the distance to the system (Dupuy \& Kraus 2013) and the radius of each component based on evolutionary models (Saumon \& Marley 2008), we can determine a probable range of physical properties for the binary.  The effective temperature of the primary is 550 -- 600~K, and that of the secondary is 450~K. Temperatures warmer or cooler by 50~K can be excluded as they result in significant discrepancies in brightness between the observed and calculated spectra.  
The shapes of the spectral distributions show that the atmospheres of both components have either very thin or no chloride/sulfide cloud layers, with a sedimentation parameter $f_{\rm sed} \gtrsim 6$. We find that the masses of the primary and secondary are around 30 and $22\,M_{\rm Jup}$, 
respectively, and that the age of the system is 4 -- 8 Gyr. This age is consistent with astrometric measurements by Dupuy \& Kraus (2013) which show that the system has kinematics  intermediate between the thin and thick disk  populations 
of the Galaxy. The system may be metal poor based on the $H - K$ colors of both components, which would also generally be consistent with an older age. 

Coeval binary systems such as  WISEPC J121756.91$+$162640.2AB offer a powerful probe of the atmospheric changes that occur at very low temperatures. At  $T_{\rm eff} \approx 500$~K, the alkali elements are condensing, and cloud decks of sulfides and chlorides form. This system offers an insight into the interplay between temperature, gravity, metallicity and cloud formation in cold atmospheres, and will provide a benchmark for the models as they 
are improved with more complete treatment of opacities, clouds and turbulence. In the near term, model atmospheres with a range of metallicity and mixing efficiency would enable a significant improvement in our understanding of the recently discovered Y dwarf population.

\acknowledgments

This research was supported by NSF grants AST09-09222
awarded to MCL. DS is supported by NASA Astrophysics Theory grant NNH11AQ54I.
Based on observations obtained at the Gemini Observatory, which is operated by the Association of Universities for
 Research in Astronomy, Inc., under a cooperative agreement with the
    NSF on behalf of the Gemini partnership: the National Science
    Foundation (United States), the Science and Technology Facilities
    Council (United Kingdom), the National Research Council (Canada),
    CONICYT (Chile), the Australian Research Council (Australia),
    Minist\'{e}rio da Ci\^{e}ncia, Tecnologia e Inova\c{c}\~{a}o (Brazil)
    and Ministerio de Ciencia, Tecnolog\'{i}a e Innovaci\'{o}n Productiva
    (Argentina). SKL's research is supported by Gemini Observatory.  This publication makes use of data products from the Wide-field Infrared Survey Explorer, which is a joint project of the University of California, Los Angeles, and the Jet Propulsion Laboratory/California Institute of Technology, funded by the National Aeronautics and Space Administration. This research has made use of the NASA/ IPAC Infrared Science Archive, which is operated by the Jet Propulsion Laboratory, California Institute of Technology, under contract with the National Aeronautics and Space Administration.

%% The reference list follows the main body and any appendices.
%% Use LaTeX's thebibliography environment to mark up your reference list.
%% Note \begin{thebibliography} is followed by an empty set of
%% curly braces.  If you forget this, LaTeX will generate the error
%% "Perhaps a missing \item?".
%%
%% thebibliography produces citations in the text using \bibitem-\cite
%% cross-referencing. Each reference is preceded by a
%% \bibitem command that defines in curly braces the KEY that corresponds
%% to the KEY in the \cite commands (see the first section above).
%% Make sure that you provide a unique KEY for every \bibitem or else the
%% paper will not LaTeX. The square brackets should contain
%% the citation text that LaTeX will insert in
%% place of the \cite commands.

%% Note that the style of the \bibitem labels (in []) is slightly
%% different from previous examples.  The natbib system solves a host
%% of citation expression problems, but it is necessary to clearly
%% delimit the year from the author name used in the citation.
%% See the natbib documentation for more details and options.

\clearpage

\begin{figure}
\includegraphics[angle=-90,scale=.7]{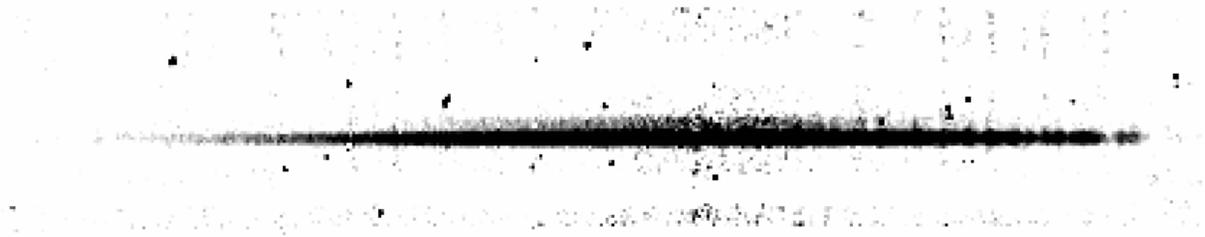}
\caption{The flat fielded, sky-subtracted and rectified  $J$-band spectral order of  WISE 1217$+$16AB from 2013 May 8. The faint, more peaked in wavelength, trace of the cooler component can be seen above the primary in this image. The wavelength range is approximately 1.18 to 1.34 $\mu$m, from left to right.
\label{fig1}}
\end{figure}

\begin{figure}
\includegraphics[angle=-90,scale=.65]{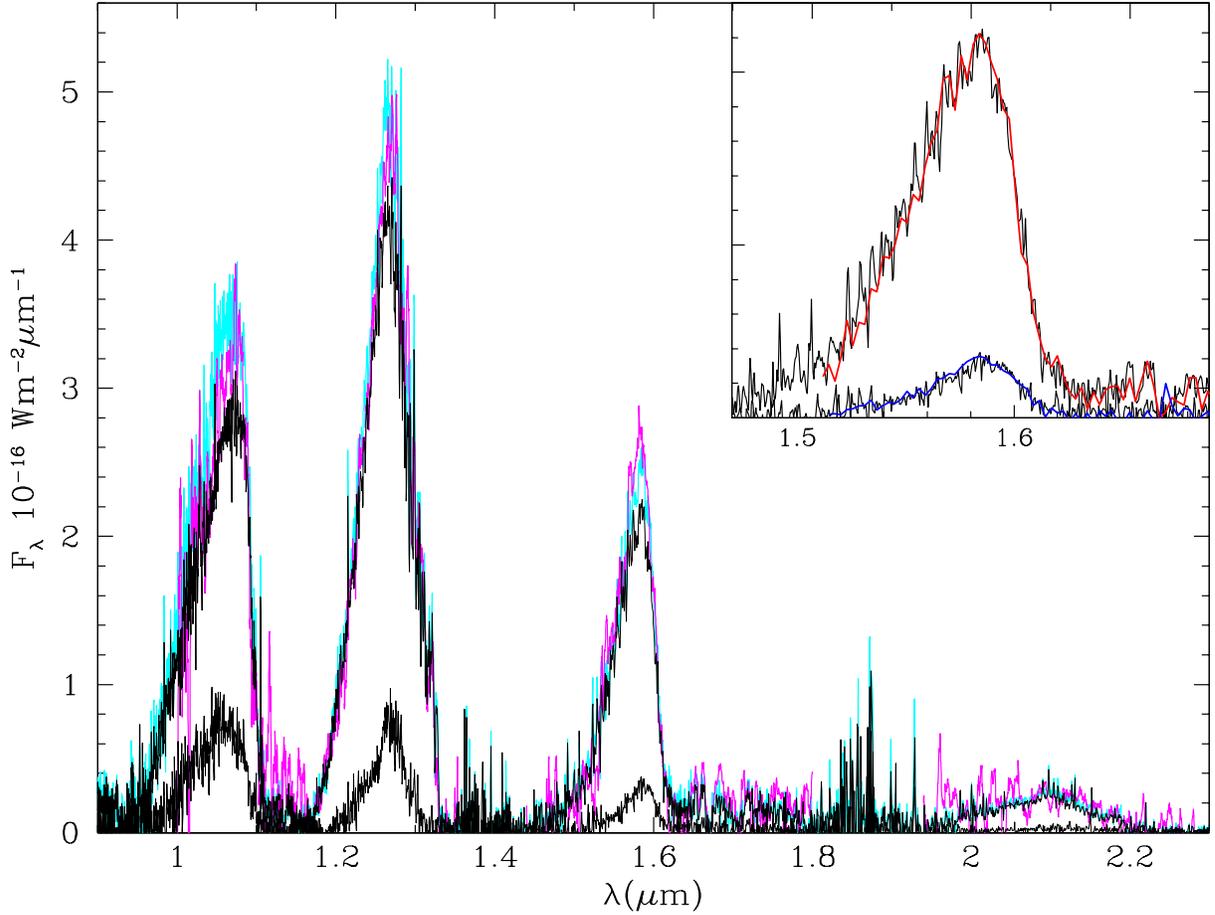}
\caption{
The complete one-dimensional GNIRS spectrum of each component of the  WISE 1217$+$16 binary is shown in black. Also shown is the summed spectrum (cyan) and an observed unresolved spectrum  from Kirkpatrick et al. (2011, magenta). The inset plot compares our $H$-band spectral order to the resolved spectra obtained previously at Keck by Liu12 (red and blue lines).
\label{fig2}}
\end{figure}

\begin{figure}
\includegraphics[angle=0,scale=.75]{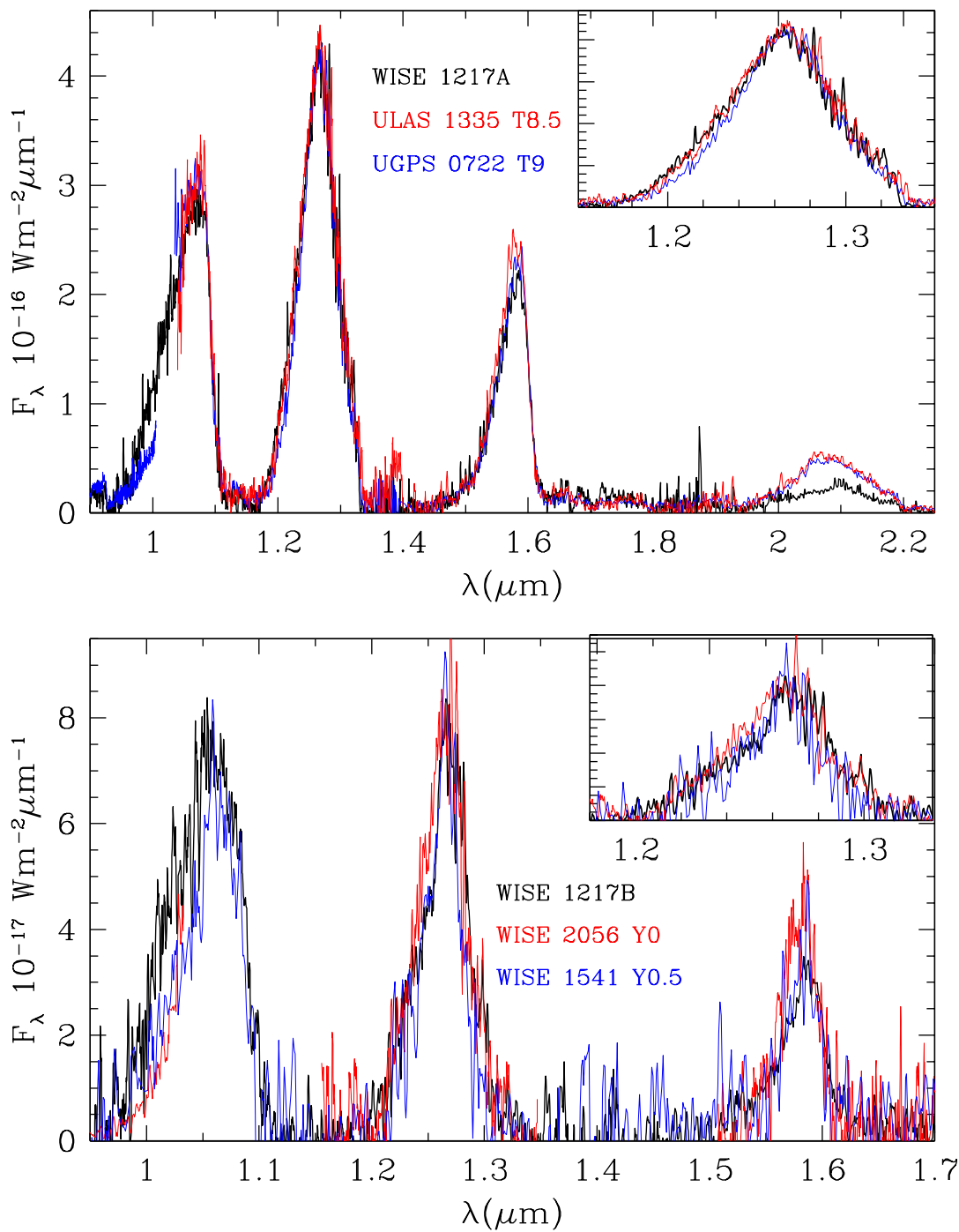}
\caption{The spectrum of each component of the  WISE 1217$+$16 binary (black line) is compared to template T and Y types (red and blue lines). The comparison spectra have been scaled to the $J$-band flux peaks. The inset plots zoom in on the $J$-band flux peak, the width of which is diagnostic of type. Note that the two spectra are plotted on different flux scales and on different wavelength scales.
\label{fig3}}
\end{figure}

\begin{figure}
\includegraphics[angle=0,scale=.65]{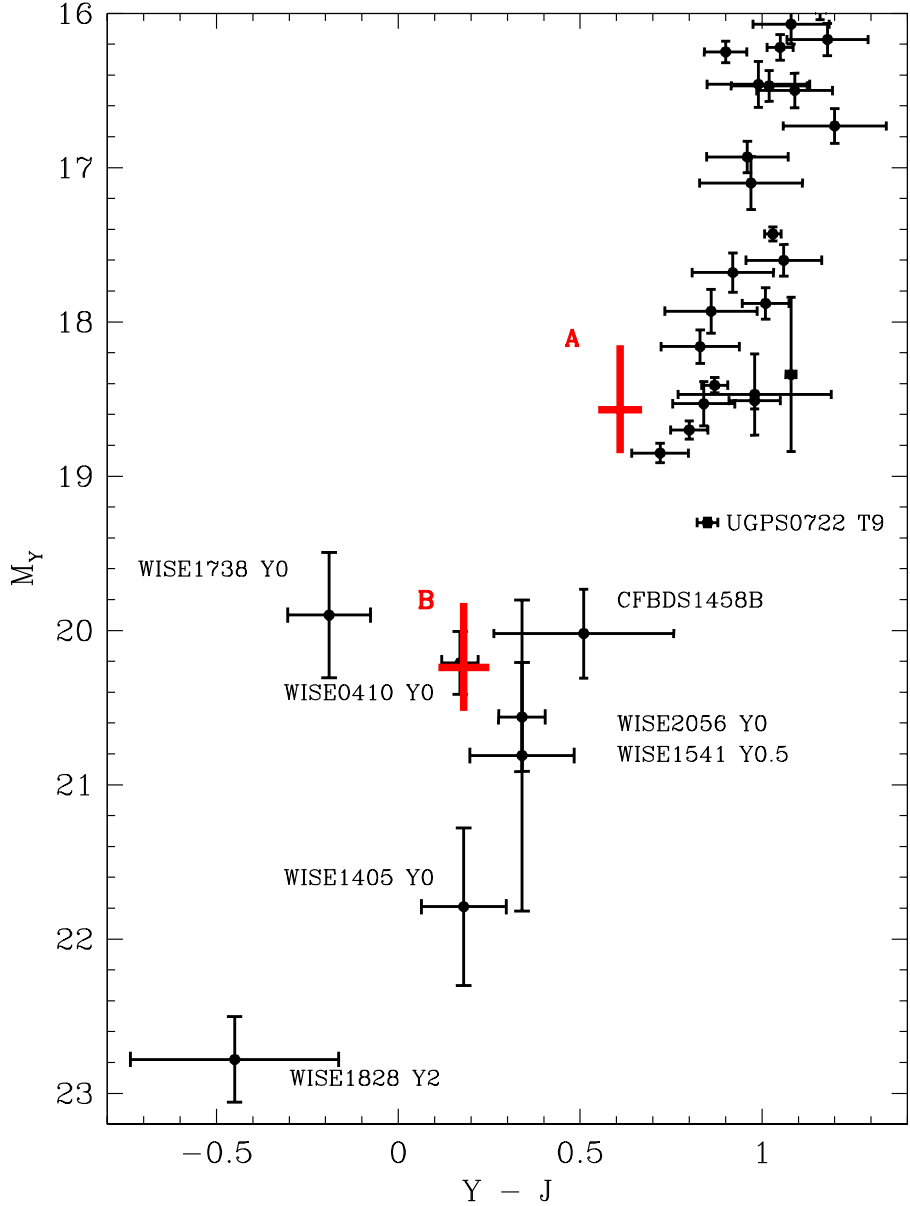}
\caption{The location of  WISE 1217$+$16A and  WISE 1217$+$16B in the {$Y - J$,$M_Y$} 
 color-magnitude diagram. The latest T-types and Y dwarfs are identified (discovery references are given in the caption for Figure 5). No spectral type is given for CFBDS1458B as no resolved near-infrared spectrum is published. All data are on the  Mauna Kea Observatories (MKO) system (Tokunaga \& Vacca 2005). 
\label{fig4}}
\end{figure}

\begin{figure}
\includegraphics[angle=-90,scale=.6]{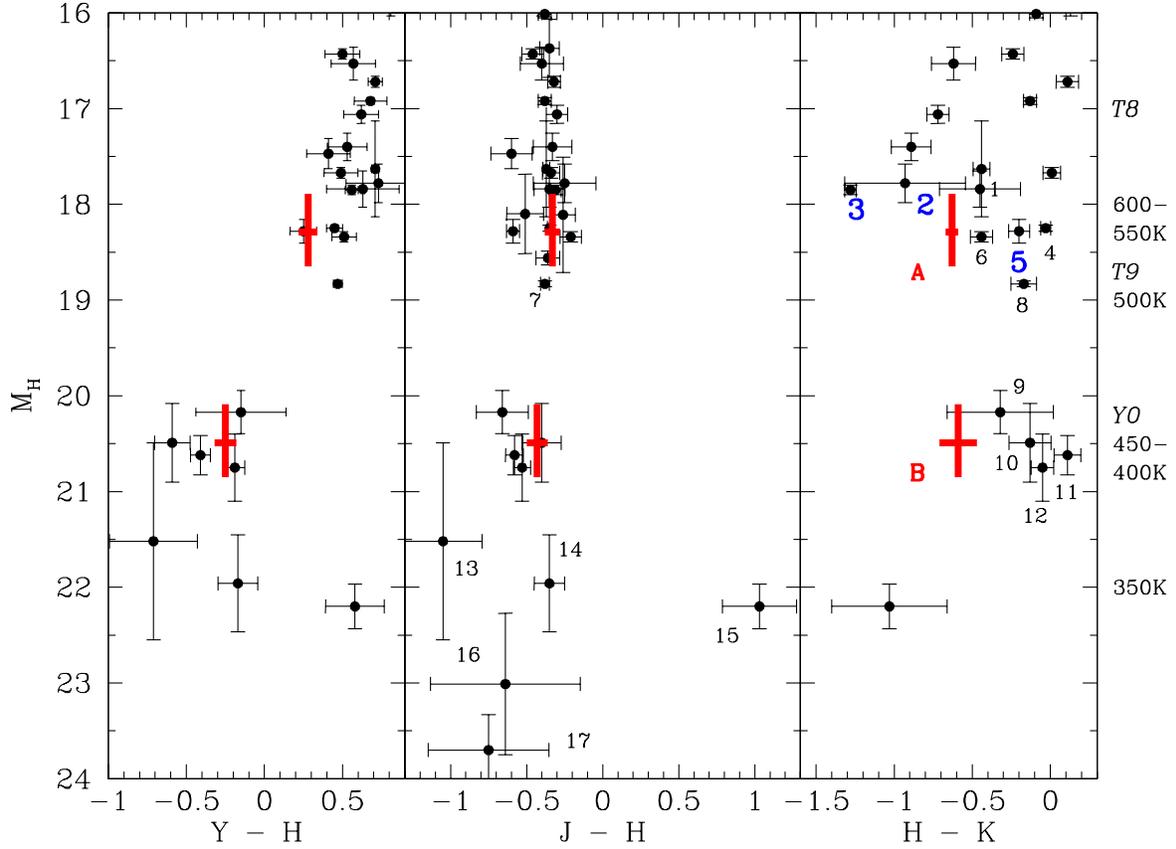}
\caption{
The location of  WISE 1217$+$16A and  WISE 1217$+$16B in near-infrared color-magnitude diagrams is indicated. 
Sources labelled 2, 3 and 5 are discussed in the text.
The typical spectral types and derived  $T_{\rm eff}$ 
at particular $M_H$ are shown on the right axis. All data are on the MKO 
photometric system. Sources are: (1) CFBDSIR J145829+101343A  (Delorme et al. 2010, Liu et al. 2011)
(2) BD $+01^{\circ} 2920$B (Pinfield et al. 2012) 
(3) SDSS J141624.08+134826.7B (Burgasser, Looper \& Rayner 2010; Burningham et al. 2010; Scholz 2010) 
(4) ULAS J133553.45+113005.2 (Burningham et al. 2008)
(5) Wolf 940B (Burningham et al. 2009) 
(6) CFBDS J005910.90-011401.3 (Delfosse et al. 2008)
(7) $\xi$ UMaC (Wright et al. 2013)
(8) UGPS J072227.51−054031.2 (Lucas et al. 2010)
(9)  CFBDSIR J145829+101343B  (Delorme et al. 2010, Liu et al. 2011)
(10)  WISEP J173835.52+273258.9   (Cushing et al. 2011)
(11) WISEPC J014807.25–720258.8 (Cushing et al. 2011)
 (12) WISEPC J205628.90+145953.3 (Cushing et al. 2011)
(13) WISEP J154151.65–225025.2 (Cushing et al. 2011)
(14) WISEPC J140518.40+553421.5 (Cushing et al. 2011)
(15) WISEP J182831.08+265037.8     (Cushing et al. 2011)
(16) WISE J035934.06–540154.6 (Kirkpatrick et al. 2012)
(17) WISE J064723.23−623235.5 (Kirkpatrick et al. 2013).
\label{fig5}}
\end{figure}

\begin{figure}
\includegraphics[angle=0,scale=.7]{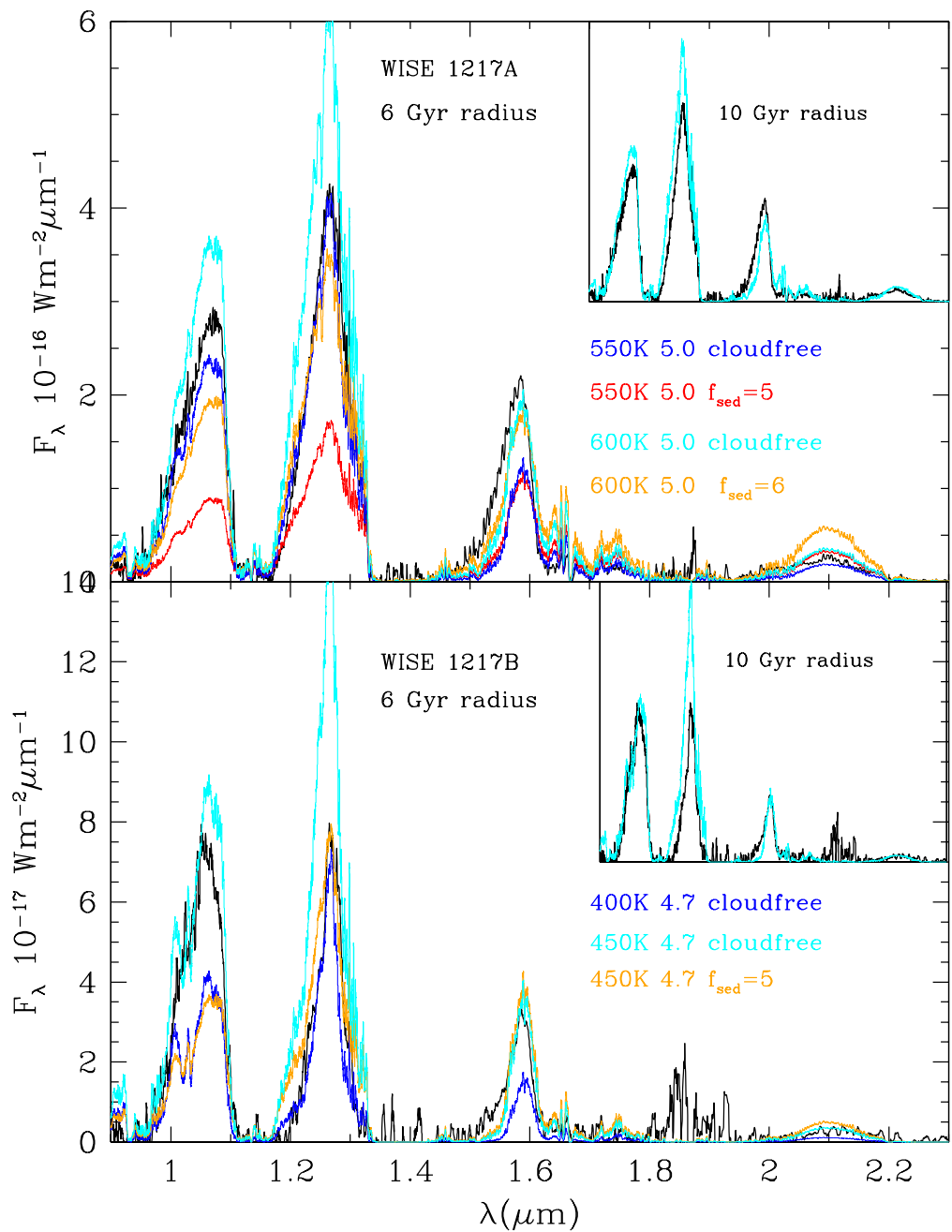}
\caption{The spectrum of each component of WISE 1217+16 (black curve) is compared to synthetic spectra generated by the Morley et al. (2012) models. Legends give each model's  $T_{\rm eff}$, log $g$ and cloudiness parameter.
The synthetic spectra have been scaled by the known distance to the binary (Table 1), and the radius of each component as given by the evolutionary models for an age of 6 Gyr (Table 2). The inset shows a comparison using fluxes scaled by the radii corresponding to an age of 10 Gyr. 
\label{fig6}}
\end{figure}

%% Tables should be submitted one per page, so put a \clearpage before
%% each one.

%% Two options are available to the author for producing tables:  the
%% deluxetable environment provided by the AASTeX package or the LaTeX
%% table environment.  Use of deluxetable is preferred.
%%

%% Three table samples follow, two marked up in the deluxetable environment,
%% one marked up as a LaTeX table.

%% In this first example, note that the \tabletypesize{}
%% command has been used to reduce the font size of the table.
%% Note also that the \label command needs to be placed 
%% inside the \tablecaption.

\clearpage

\begin{deluxetable}{llll}
\tablewidth{0pt}
\tablecaption{Observational Properties of  WISEPC J121756.91$+$162640.2AB}
\tablehead{
\colhead{Property} & \colhead{Component A} & \colhead{Component B} & \colhead{Reference}\\
}
\startdata
Parallax (mas) &  \multicolumn{2}{c}{99 $\pm$ 16} & Dupuy  \& Kraus 2013 \\
Proper motion RA (mas/yr) & \multicolumn{2}{c}{786 $\pm$ 42} & Dupuy \& Kraus. 2013 \\
Proper motion Decl. (mas/yr) & \multicolumn{2}{c}{$-$1224 $\pm$ 27} & Dupuy \& Kraus 2013 \\
Separation (arcseconds) &   \multicolumn{2}{c}{$0\farcs 76$} & {Liu et al.  2012}\\
$Y_{\rm MKO}$ &\multicolumn{2}{c}{18.38  $\pm$ 0.04} & {Liu et al.  2012}\\
$J_{\rm MKO}$ & \multicolumn{2}{c}{17.83  $\pm$ 0.02} & {Kirkpatrick et al. 2011}\\
$H_{\rm MKO}$ &\multicolumn{2}{c}{18.18  $\pm$ 0.05} & {Kirkpatrick et al. 2011}\\
$K_{\rm MKO}$ &\multicolumn{2}{c}{18.80  $\pm$ 0.04} & {Liu et al.  2012}\\
$W1_{WISE}$ &  \multicolumn{2}{c}{16.80 $\pm$ 0.13} & {\it WISE} All Sky Release \\
$W2_{WISE}$ &  \multicolumn{2}{c}{13.09 $\pm$ 0.03} & {\it WISE} All Sky Release \\
$W3_{WISE}$ &  \multicolumn{2}{c}{12.12 $\pm$ 0.31} & {\it WISE} All Sky Release \\
$3.6_{Spitzer}$ &  \multicolumn{2}{c}{15.44 $\pm$ 0.02} &  {Kirkpatrick et al. 2011}\\
$4.5_{Spitzer}$ &  \multicolumn{2}{c}{13.11 $\pm$ 0.02} &  {Kirkpatrick et al. 2011}\\
Spectral Type & T8.5 & Y0--0.5 & This paper \\
$Y_{\rm MKO}$ & 18.59 $\pm$ 0.04 & 20.26 $\pm$ 0.04  & {Liu et al.  2012}\\
$J_{\rm MKO}$ & 17.98 $\pm$ 0.02 & 20.08 $\pm$ 0.03  & {Liu et al.  2012}\\
$H_{\rm MKO}$ & 18.31 $\pm$ 0.05 & 20.51 $\pm$ 0.06  & {Liu et al.  2012}\\
$K_{\rm MKO}$ & 18.94 $\pm$ 0.04 & 21.10 $\pm$ 0.12  & {Liu et al.  2012}\\
\enddata
\end{deluxetable}

\begin{deluxetable}{lcccc}
\tablewidth{0pt}
\tablecaption{Evolutionary Parameters for $T_{\rm eff}$(A)= 550 -- 600$\,$K and  $T_{\rm eff}$(B)= 400 -- 450$\,$K.}
\tablehead{
\colhead{Property} & \multicolumn{2}{c}{Primary} & \multicolumn{2}{c}{Secondary}\\
 & {550 K} &  \colhead{600 K} & \colhead{400 K}  & \colhead{450 K}\\
}
\startdata
& \multicolumn{3}{c}{\bf 4 Gyr}& \\
Mass ($M_{\rm Jup}$) & 25 & 26 & 16 & 19 \\
Radius ($R_{\odot}$) & 0.0953 & 0.0936 & 0.1016 & 0.0993 \\
log $g$ (cm s$^{-2}$) & 4.859 & 4.929 &  4.600 & 4.699 \\
&\multicolumn{3}{c}{\bf 6 Gyr}& \\
Mass ($M_{\rm Jup}$) &  29 & 34 & 19  &  22 \\
Radius ($R_{\odot}$) &  0.0917 & 0.0895 &  0.0982 & 0.0964 \\
log $g$ (cm s$^{-2}$) & 4.966 & 5.053   &  4.708 & 4.757 \\
&\multicolumn{3}{c}{\bf 8 Gyr}& \\
Mass ($M_{\rm Jup}$) &  33 & 37 &  21 & 25 \\
Radius ($R_{\odot}$) &  0.0891 & 0.0871 &  0.0961 & 0.0934 \\
log $g$ (cm s$^{-2}$) &  5.044 & 5.119 & 4.773 &  4.877 \\
&\multicolumn{3}{c}{\bf 10 Gyr}& \\
Mass ($M_{\rm Jup}$) &  35 & 42 &  23 & 28 \\
Radius ($R_{\odot}$) &  0.0870 & 0.0851 &  0.0942 & 0.0915 \\
log $g$ (cm s$^{-2}$) &  5.108 & 5.180 & 4.832 &  4.977 \\
\enddata
\end{deluxetable}

\begin{deluxetable}{lcc}
\tablewidth{0pt}
\tablecaption{Physical Properties of WISE 1217$+$16A and  WISE 1217$+$16B}
\tablehead{
\colhead{Property} & \colhead{Component A} & \colhead{Component B} \\
}
\startdata
Age (Gyr)&\multicolumn{2}{c}{4 -- 8} \\
$T_{\rm eff}$ (K) & 550 -- 600 & 450 \\
Mass ($M_{\rm Jup}$) & 30 $\pm$ 5 & 22  $\pm$ 2 \\
Radius ($R_{\odot}$) & 0.091 $\pm$ 0.004 & 0.096 $\pm$ 0.003 \\
log $g$ (cm s$^{-2}$) & 5.0 $\pm$ 0.1 &  4.8 $\pm$ 0.1 \\
\enddata
\end{deluxetable}

\end{document}